\documentclass[pra, twocolumn]{revtex4}
\usepackage{graphicx}
\begin{document}
\newcommand{\e}{\mathrm{e}}
\newcommand{\x}{{\bf r}}
\newcommand{\K}{{\bf k}}
\newcommand{\y}{{\bf y}}
\newcommand{\D}{{\rm d}}
\newcommand{\p}{{\bf p}}
\newcommand{\tr}{\mathrm {Tr}}
\newcommand{\de}{:=}

\title{Elastic scattering losses from colliding BEC's}

\author{Pawe{\l} Zi\'{n}$^1$, Jan Chwede\'{n}czuk$^1$  and Marek Trippenbach$^{1,2}$}

\affiliation{$^1$ Institute for Theoretical Physics, Warsaw
University, Ho\.{z}a 69, PL-00-681 Warsaw, Poland, \\ $^{2}$ Soltan
Institute for Nuclear Studies, Ho\.{z}a 69, PL-00-681 Warsaw,
Poland.}

\begin{abstract}
Bragg diffraction divides a Bose-Einstein condensate into two
overlapping components, moving with respect to each other with high
momentum. Elastic collisions between atoms from distinct wave
packets can significantly deplete the condensate. Recently Zi\'{n}
{\it et al.} (Phys. Rev. Lett. {\bf 94}, 200401 (2005)) have
introduced a model of two counter-propagating atomic Gaussian
wavepackets incorporating dynamics of the incoherent scattering
processes. Here we study the properties of this model in detail,
including the nature of the transition from spontaneous to
stimulated scattering. Within the first order approximation we
derive analytical expressions for the density matrix and anomalous
density which provides excellent insight into correlation properties
of scattered atoms.
\end{abstract}

\maketitle

\section{Introduction}

A light-induced potential applied to a Bose-Einstein condensate can
be used to make high momentum daughter BEC wavepackets, which
propagate through the parent condensate
\cite{Ovchinnikov99,Kozuma99,Stenger}. Such technique have been used
to make an atom laser \cite{nist_al}, to study the coherence
properties of condensates \cite{Stenger,nist_cl_exp,cl_theory}, and
to study nonlinear four-wave mixing (4WM) of coherent matter waves
\cite{Meystre,4WM_1,Tripp,4WM_2}. In the process of a collision
between two condensates, inevitably, some atoms would scatter away,
forming often a noticeable hallo around the region of collision. The
hallo at the level above noise indicates a profuse elastic
scattering losses, and has been observed in various experiments, for
example \cite{Kozuma99,johny,Katz}.

\begin{figure}[htb]
\centering
\includegraphics[scale=0.3, angle=0]{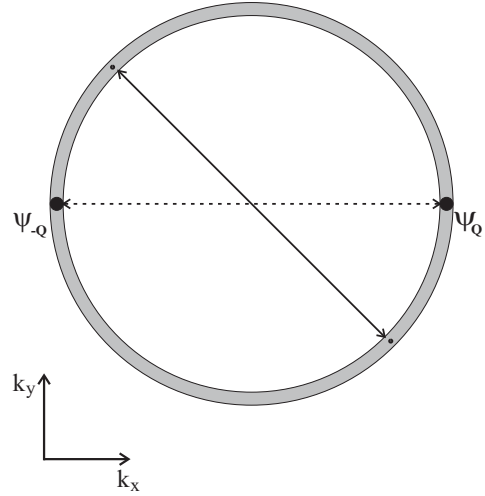}
\caption{Collision of condensates in momentum space. Big dots
denoted as  $\psi_{\pm Q}$ represent colliding condensates. Small
dots represent scattered atoms. Dashed lines represent anihilation
of atoms,  solid lines represent creation of atoms; it should help
to imagine anihilation of two atoms, one from each of the
condensates, and creation of two atoms in such states, that the
momentum and energy of the system is conserved.} \label{fig1}
\end{figure}

A scheme of such a process is illustrated in Fig.~\ref{fig1}, where
the collision of two condensates having mean wavevector $\pm Q$ is
shown in the momentum space. The condensates are denoted by large
dots and marked as $\psi_{\pm Q}$. Atoms from these two
counter--propagating wavepackets undergo elastic collisions and can
be scattered out into all the modes permitted by the energy and
momentum conservation (a pair of such states is marked with small
dots). These modes lie within the three dimensional shell, which in
Fig.~\ref{fig1} is represented by the gray ring.

The elastic scattering loss is not accounted for by the
Gross-Pitaevskii Equation (GPE). Recently, some modifications of the
GPE were proposed in order to incorporate this loss into the mean
field dynamics. This has been done either within the slowly varying
envelope approximation by adding an imaginary part to the scattering
length \cite{Band}, or by including a stochastic component to mimic
quantum field fluctuations \cite{Chwed,norrie}. Interesting results were
also obtained using method based on the field theory formulation in
the lowest order of the perturbation approximation \cite{Bach,Yuro}.
So as to test the validity of various approximate schemes an exact
(nonperturbative) model of collisional losses was proposed
\cite{Zin}. It is valid both in the regime of spontaneous initiation
and Bose enhancement. This model assumes spherical non-spreading
Gaussians for the colliding wave packets and is capable of treating
the number of scattered atoms as well as their statistical
properties through the higher order correlation functions.

Here we study this model in more detail, including the nature of the
transition from spontaneous to stimulated scattering and statistical
properties of scattered atoms. We also provide a link to previous
results obtained within the perturbation approximation.

\section{The model}

\begin{figure}[htb]
\centering
\includegraphics[scale=0.3, angle=0]{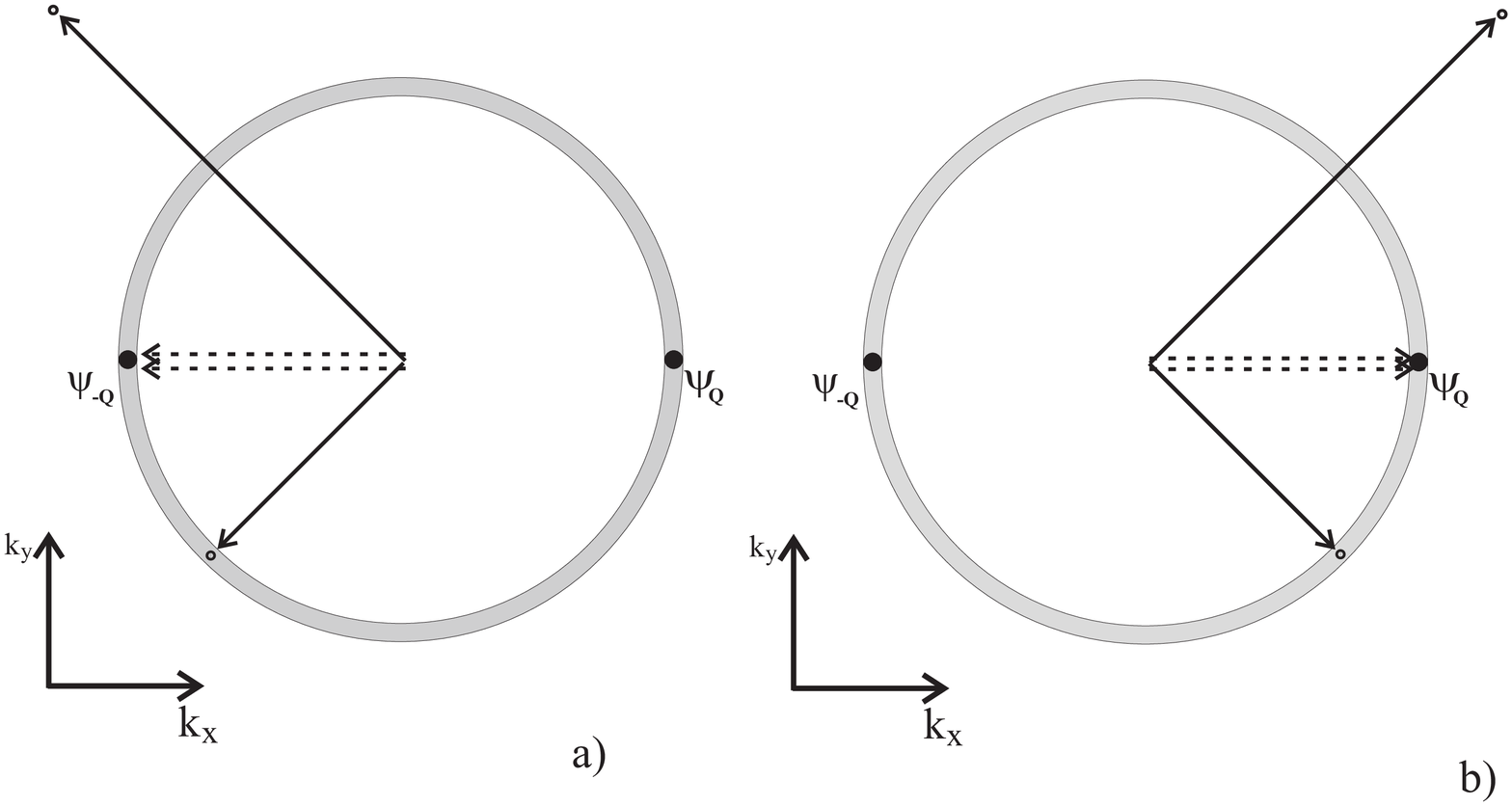}
\caption{Visualisations of
$g\hat\delta^\dagger(\x,t)\left[\psi^2_Q(\x, t)+\psi^2_{-Q}(\x,
t)\right]$term.  Figure a$)$ represents the term proportional to
$\psi^2_{-Q}$. Two atoms are anihilated from $\psi_{-Q}$ and
scattered to such states that the momentum is conserved. Analogeous
plot b$)$ is a visualisation of term proportional to
$\psi^2_Q$.}\label{fig2}
\end{figure}

A system of Bosons interacting via contact potential is described by the Hamiltonian
\begin{eqnarray}
\label{ham1} &&H=-\int d^3 r \, \hat \Psi^\dagger(\x,t)
\frac{\hbar^2\nabla^2}{2m}\hat\Psi(\x,t)\nonumber\\
&&+\frac{g}{2}\int \mbox{d}^3 r \, \hat{\Psi}^\dagger({\bf
r},t)\hat{\Psi}^\dagger(\x,t) \hat{\Psi}(\x,t) \hat{\Psi}({\bf
r},t),
\end{eqnarray}
where $\hat\Psi(\x,t)$ is a field operator satisfying equal time
bosonic commutation relations, $m$ is the atomic mass and $g$
determines the strength of the inter-atomic interactions. Since the
Hamiltonian (\ref{ham1}) is of the fourth order in $\hat\Psi$, the
Heisenberg equation governing the evolution of the field,
\begin{eqnarray}
\label{full_evolve}i\hbar\partial_t\hat{\Psi}({\bf
r},t)=-\frac{\hbar^2\nabla^2}{2m}\hat{\Psi}({\bf
r},t)+g\hat{\Psi}^\dagger(\x,t) \hat{\Psi}(\x,t) \hat{\Psi}({\bf
r},t),
\end{eqnarray}
is nonlinear and thus, in general, analytically and numerically
untractable.  However, for some physical systems, a Bogoliubov
approximation can be applied leading to linear Heisenberg equations.
The idea underlying this approximation states that for some cases
the field operator might be split into two parts: $\psi$ and
$\hat\delta$. The first contribution describes macroscopically
occupied modes, where the fluctuations are usually small; hence its
operator character might be dropped ($\psi$ becomes a c-number
wave-function satisfying GPE). The second part $\hat\delta$, called,
``above mean field part'', representing fluctuations, requires full
quantum mechanical treatment. Since the initial state of the system
consists of two counter-propagating atomic wave-packets and the
``sea'' of unoccupied modes, Bogoliubov approximation can be
applied. The splitting of the bosonic field is performed  in the
following manner:

\begin{eqnarray}
\hat\Psi(\x, t)&=&\psi_Q(\x, t)+\psi_{-Q}(\x, t)+\hat{\delta}(\x,
t), \label{deco1}
\end{eqnarray}
where the subscript $\pm Q$ denotes the mean momentum of the
colliding condensates and $\psi_Q(\x, t)+\psi_{-Q}(\x, t)$ satisfies
the time-dependent GPE. Upon inserting Eq.~(\ref{deco1}) into the
Heisenberg equation (\ref{full_evolve}) one obtains on the right
hand side several terms, of which, in the spirit of the Bogoliubov
approximation, we keep only those up to the first order in
$\hat\delta$,
\begin{eqnarray}
\label{Full}&&i\hbar\partial_t\hat\delta(\x, t)= \\
&&\left[-\frac{\hbar^2\nabla^2}{2m}+2g|\psi_Q(\x,
t)|^2+2g|\psi_{-Q}(\x, t)|^2 \right. \nonumber \\
 && \left.\phantom{\frac{.}{.}}+2g\psi^*_Q(\x, t)\psi_{-Q}(\x, t)+2g\psi^*_{-Q}(\x,
t)\psi_Q(\x, t) \right] \hat\delta(\x, t)\nonumber\\
\nonumber&&+g\left[2\psi_Q(\x, t)\psi_{-Q}(\x, t)+\psi^2_Q(\x,
t)+\psi^2_{-Q}(\x, t)\right]\hat{\delta}^\dagger(\x, t).
\end{eqnarray}
The above equation can be simplified using following arguments.
First, we expect that for as long as the mean kinetic energy of the
scattered atoms ($\hbar^2Q^2/(2m)$) is much larger than the
interaction energy ($gn$, where $n$ is the mean density of the atoms
in the condensate), the mean field energy terms $[2g|\psi_Q(\x,
t)|^2+2g|\psi_{-Q}(\x,t)|^2+2g\psi^*_Q(\x, t)\psi_{-Q}(\x,
t)+2g\psi^*_{-Q}(\x, t)\psi_Q(\x, t)]$, can be dropped, leading to
\begin{eqnarray}
\label{Full1}&&i\hbar\partial_t\hat\delta(\x, t)=-
\frac{\hbar^2\nabla^2}{2m}\hat\delta(\x, t) \\
&&+g\left[2\psi_Q(\x, t)\psi_{-Q}(\x, t)+\psi^2_Q(\x,
t)+\psi^2_{-Q}(\x, t)\right]\hat{\delta}^\dagger(\x, t). \nonumber
\end{eqnarray}
To make further simplifications more transparent, we visualize
effects of various terms in the momentum space using schematic
picture. The first term, $2g\hat\delta^\dagger(\x,t)\psi_Q(\x,
t)\psi_{-Q}(\x, t)$, is a source term. It corresponds to
annihilation of two particles from counter-propagating condensates
and creation of two particles in the field of the ``sea'' of
scattered atoms. An example of such a process is illustrated in
Fig.~\ref{fig1}. The second term, $
g\hat\delta^\dagger(\x,t)\left[\psi^2_Q(\x, t)+\psi^2_{-Q}(\x,
t)\right]$, as shown in Fig.~\ref{fig2}, describes annihilation of a
pair of particles from the same condensates (either $\psi_Q$ or
$\psi_{- Q}$) and creation of two particles in the $\hat\delta$
field. As seen from Fig.~\ref{fig2}, this process is non-resonant,
leading to violation of the conservation laws.

%The third term,
%\begin{eqnarray}
%\label{a2}a_3=2g\hat\delta(\x,t)\left[\psi^*_Q(\x, t)
%\psi_{-Q}(\x, t)+\psi^*_{-Q}(\x, t)\psi_Q(\x, t)\right],
%\end{eqnarray}
%describes the annihilation of a pair of particles; one from the
%condensate and the other from the $\hat\delta$ field, followed by
%the creation of a pair of particles; one in a condensate with a
%opposite momentum and the second one in the $\hat\delta$ field. As
%demonstrated in Fig.~\ref{fig3}, this term is also of-resonant.
%\begin{figure}[htb]
%\centering
%\includegraphics[scale=0.3, angle=0]{fig4.eps}
%\caption{Visualisations of both terms in Eq.(\ref{a2}). Figure a$)$
%represents the term proportional to $\psi^*_{-Q}\psi_{-Q}$. An atom
%from the $\hat\delta$ collides with an atom from the $\psi_{-Q}$
%condensate leading to the scattering one of them to $\psi_Q$
%condensate and another one back to the $\hat\delta$ field in such a
%way that the momentum is conserved. Analogeous plot b$)$ is a
%visualisation of term proportional to $\psi^*_{-Q}(\x, t)\psi_Q(\x,
%t)$.} \label{fig3}
%\end{figure}
The above arguments imply that the Bogoliubov equation (\ref{Full1})
can be simplified by dropping the off-resonant term and leaving only
the source term. The evolution equation for the $\hat\delta$ field
finally becomes
\begin{equation}
\label{heis}i\hbar\partial_t\hat\delta(\x, t)= -\frac{\hbar^2\nabla^2}{2m}
\hat\delta(\x, t)+2g\psi_Q(\x,
t)\psi_{-Q}(\x, t)\hat{\delta}^\dagger(\x, t).
\end{equation}
%One can argue that such an approximation gives correct results if
%and only if the kinetic energy associated with the center-of-mass
%motion is much larger than the interaction energy per particle,
%$\hbar^2Q^2/(2m) \gg gn$, where n is the average density of the
%particles in the condensates. Numerical proof of the above statement
%in the simplest case of two plane matter waves was given in
%\cite{Bach}.

To simplify the dynamics even further we define and compare three
characteristic timescales of the problem. Let $\sigma$ and $N/2$ be
the width and number of particles in each of the colliding
condensates. Then, the collisional time, (the time it takes for each
wave-packet to pass through its colliding partner) is defined by:
$t_{C}=(m\sigma)/(\hbar Q)$. Another characteristic timescale of the
problem is the linear dispersion time (time of the spread of the
wave-packet due to kinetic energy term), $t_{LD}=m\sigma^2/\hbar$
\cite{Tripp}. In the same manner the nonlinear dispersion time,
which is the time of ballistic expansion in Thomas Fermi
approximation \cite{CDG}, can be defined:  $t_{ND}=\sqrt{\pi^{3/2}
m\sigma^5/gN}$.
%and the nonlinear time - characteristic time for nonlinear interactions -
%$t_{NL}=(\hbar\sigma^3)/(gN)$.
The dynamics of the system depends on the relations between these
timescales. It is convenient to introduce the following
dimensionless parameters: $t_{LD}/t_C=\beta$ and
$(t_{LD}/t_{ND})^2=\alpha$. When the number of elastically scattered
atoms is small in comparison with the total number of atoms in both
wave-packets and both linear and nonlinear dispersion timescales are
much longer than the collision time ($(t_{LD}/t_C)=\beta \gg 1$ and
$(t_{ND}/t_C)^2=\beta^2/\alpha\gg 1$), we can neglect the change in
population and shape of the macroscopically occupied functions
$\psi_{\pm Q}({\x},t)$ during the collision. In our model we assume
Gaussian shape of the colliding condensates
\begin{eqnarray}
\label{cond} &&\psi_{\pm Q}(\x,t) =
\sqrt{\frac{N}{2\pi^{3/2}\sigma^{3}}} \exp \left[ \pm iQx_1
-\frac{i\hbar t
Q^2}{2m} \right]\times\nonumber\\
&&\exp\left[-\frac{1}{2\sigma^2}\left(\left(x_1\mp\frac{\hbar Qt}{m}\right)^2
+x_2^2+x_3^2\right)\right],
\end{eqnarray}
where $\x=(x_1,x_2,x_3)$. Although the assumption that the Gaussians
have spherical symmetry is not the most general, it makes the
problem numerically tractable and allows to get insight into all
relevant physical aspects of the system.

Upon substitution (\ref{cond}) into (\ref{heis}), and then rescaling
the variables $\x/\sigma \rightarrow \x $ we obtain:
\begin{equation}
i\beta\partial_t\hat{\delta}(\x,t)=
-\frac{1}{2}\Delta \hat{\delta}(\x,t) +\alpha e^{-r^2-t^2- i \beta
t}\hat{\delta}^\dagger(\x,t).\label{EofM}
\end{equation}
Next we switch to dimensionless field operators substituting
 $\hat{\delta}(\x,t)\sigma^{3/2} \rightarrow
\hat{\delta}(\x,t)$. Notice, that assumption of spherical symmetry
of the colliding wave-packets imposes such symmetry on
Eq.~(\ref{EofM}). Thus it is convenient to decompose $\hat{\delta}$
into the basis of spherical harmonics
\begin{eqnarray}
\label{decomp}\hat{\delta}(\x,t)&=&\sum_{n,l,m}R_{n,l}(r)Y_{lm}(\theta,
\phi)\hat a_{n,l,m}(t),
\end{eqnarray}
where $\hat a_{n,l,m}$ are annihilation operators for a particle in
the mode  described by $n,l,m$ quantum numbers. There is still a
freedom of choice with regards to the set of orthogonal functions
$R_{n,l}(r)$. As we shall see below a good candidate is a set of
eigenfunctions of spherically symmetric harmonic oscillator,
\begin{equation}
R_{n,l}(r) = \sqrt{\frac{2n!a_0^{-3}}{\Gamma(l+n+\frac{3}{2})}}
\left(  \frac{r}{a_0}\right)^l
e^{-\frac{r^2}{2a_0^2}}L_n^{l+\frac{1}{2}}\left(\frac{r^2}{a_0^2}\right),
\end{equation}
where $L_n^{l+\frac{1}{2}}(x)$ is the associated Laguerre polynomial
\cite{Abram} and $a_0$, a harmonic oscillator length, is an
auxiliary free parameter that can be chosen to minimize the
computational effort. Notice, that the choice of orthogonal basis
preserves bosonic commutation relations for annihilation and
creation operators $a_{n,l,m}$.
\begin{eqnarray}
&&[a_{n,l,m},a^\dagger_{n'l'm'}]=\delta_{nn'}\delta_{mm'}\delta_{ll'} \\
&&[a_{n,l,m},a_{n'l'm'}]=0,\,\,\,\, [a^\dagger_{n,l,m},a^\dagger_{n'l'm'}]=0.
\end{eqnarray}
Substituting Eq.~(\ref{decomp}) into Eq.~(\ref{EofM}) and making use
of the completeness of the basis functions we obtain the equation
governing the evolution of operator $a_{n,l,m}$,
\begin{eqnarray}
&&i\partial_t\hat a_{n,l,m}= B_{n,l}\hat
a_{n,l,m}+D_{n,l} \hat
a_{n-1,l,m}\nonumber\\
&&+D_{n+1,l}\hat a_{n+1,l,m} +\frac{\alpha}{\beta} e^{-t^2}\sum_{n'}C_{n,n',l}
\hat a^\dagger_{n',l,-m},
\label{lsystem}
\end{eqnarray}
where $E_{n,l}=(2n+l+3/2)/a_0^2$,
$B_{n,l}=(E_{n,l}-\beta^2)/2\beta$,
$D_{n,l}=\sqrt{n(n+l+1/2)}/(2\beta a_0^2)$,
 and
\begin{eqnarray} \nonumber
&&C_{n,n',l} = \int_0^\infty r^2 \mbox{d}r \, R_{n,l}(r) \exp(-r^2)
R_{n'l}(r)=\\ \nonumber \\
\nonumber &&=\sqrt{\frac{\Gamma \left(n+l +\frac{3}{2}\right)\Gamma
\left(n' +l+\frac{3}{2}\right)}{\Gamma \left(l+\frac{3}{2}
\right)^2\Gamma \left(n+1\right)\Gamma\left(n'+1\right)}} \left(1+
a_0^2\right)^{-l-\frac{3}{2}}\\
&&\times\left[\frac{ a_0^2}{1+ a_0^2}\right]^{n+n'} F \left(
-n,-n',l+\frac{3}{2} , 1/a_0^4 \right).
\end{eqnarray}
Here $F(a,b,c,x)$ is a hypergeometric function \cite{Abram}. Notice
that all coupling coefficients are calculated analytically and the
$\hat a_{n,l,m}$ operators for different $l$ and $m$ are decoupled.
Moreover, equations (\ref{lsystem}) do not depend on quantum
number $m$. With all these simplifications the linear system of
equations (\ref{lsystem}) can be solved numerically. The general
three-dimensional problem, due to spherical symmetry simplifies to
the set of one-dimensional ordinary differential equations. In
numerical applications, for every $l$ a basis of approximately 64
modes associated with quantum number $n$ is sufficient. Moreover,
simulations show that 50 modes associated with quantum number $l$ is
usually enough. Thus the whole quantum model can be solved
numerically  on an ordinary PC within one hour of calculations!

\begin{figure}
\includegraphics[scale=0.4, angle=0]{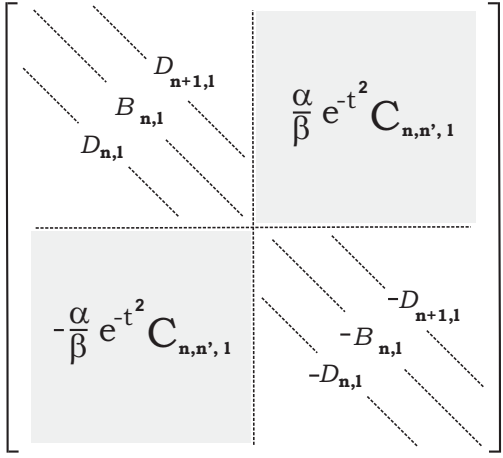}
\caption{Visualisation of ${\bf \hat A}_l(t)$ matrix from
Eq.(\ref{raw1}). It is tri-diagonal in coupling $a_{n,l,m}$'s with
$a_{n',l,m}$'s and $a^\dagger_{n,l,-m}$'s with
$a^\dagger_{n',l,-m}$'s. The coupling between $a_{n,l,m}$'s and
$a^\dagger_{n',l,-m}$'s is determined by
$-\frac{\alpha}{\beta}e^{-t^2}C_{n,n',l}$. In general, all
$C_{n,n',l}$'s are non-zero, which is represented by grey
area.}\label{matrix}
\end{figure}

We define a vector of operators:
\begin{eqnarray*}
{\bf v}_{l,m}(t) = \left( \begin{array}{l} \hat{a}_{0,l,m} \\
\hat{a}_{1,l,m} \\  \vdots
\\ \hat{a}_{n_{max},l,m} \\  \\ \hat{a}_{0,l,-m}^\dagger \\
\hat{a}_{1,l,-m}^\dagger  \\
\vdots \\ \hat{a}_{n_{max},l,-m}^\dagger \end{array} \right) ,
\end{eqnarray*}
and rewrite the equation (\ref{lsystem}) in the compact form:
\begin{eqnarray}
i\frac{\mbox{d}}{\mbox{d}t} {\bf v}_{l,m}(t) = {\bf \hat A}_l(t){\bf
v}_{l,m}(t).\label{raw1}
\end{eqnarray}
The matrix ${\bf \hat A}_l(t)$ is plotted schematically in
Fig.~\ref{matrix}.

The solution ${\bf v}_{l,m}(t)$ can be expressed in terms of the
time evolution operator:
\begin{eqnarray}
{\bf v}_{l,m}(t) = {\bf \hat U}_l(t){\bf v}_{l,m}(0).\label{row2}
\end{eqnarray}
The evolution operator satisfies equation
\begin{eqnarray}
\label{EvO}\frac{\mbox{d} {\bf \hat U}_l(t)}{\mbox{d}t} = -i{\bf \hat A}_l(t){\bf \hat U}_l(t)
\end{eqnarray}
with the initial condition ${\bf \hat U}_l(0) = {\bf 1}$. In our
calculations we evaluate ${\bf \hat A}_l(t)$ and find evolution
matrix ${\bf \hat U}_l(t)$ by solving Eq.~(\ref{EvO}). The matrix
${\bf \hat U}_l(t)$ uniquely determines full operator dynamics of
$\hat\delta(\x, t)$ and hence is sufficient to obtain any
expectation value of the variables referring to the system.

Equation (\ref{EofM}) can be treated alternatively in the Fourier
domain. Upon defining the Fourier transform in the form
\begin{eqnarray*}
\hat\delta (\x,t) = \left(\frac{\beta}{2 \pi} \right)^{3/2} \int
\mbox{d}^3 k\exp \left( i \beta {\bf k r} - i \beta k^2 t/2
\right) \hat\delta (\K,t),
\end{eqnarray*}
we can rewrite Eq.~(\ref{EofM}) as
\begin{eqnarray}
&&i\partial_t\hat{\delta}(\K,t)= \frac{\alpha\beta^2}{8 \pi^{3/2}}
e^{-t^2} \int \mbox{d}^3 k \, \exp
\left( -\frac{\beta^2 ({\bf k + k'})^2}{4}\right)\nonumber\\
&&\times \exp\left( -  i \beta t \left( 1 - \frac{k^2 +
k'^2}{2}\right) \right) \hat\delta^\dagger ({\bf
k'},t).\label{bofexp}
\end{eqnarray}
We took this particular form of the Fourier transform (with factor
$\beta$) so that the wavevector of the moving condensate is equal to
unity.

%This calculations are presented in the following chapter.
%This can be integrated analytically, using error functions:
%\begin{eqnarray}
%a_\K^{(1)} &=& -i\frac{\alpha}{\beta}
%\left(\frac{\sigma}{L}\right)^3\pi^{3/2} \frac{\sqrt{\pi}}{2} \sum_{\bf k'} \exp \left( -
%\frac{\beta^2 (\K + {\bf k'})^2}{4} - \frac{\beta^2 }{4} \left(1-\frac{k^2+k'^2}{2}
%\right)^2\right)\nonumber \\
%& & \left[ \mbox{erf} \left( t + \frac{i }{2} \beta \left( 1 - \frac{k^2+k'^2}{2}
%\right) \right) - \mbox{erf} \left( \frac{i }{2} \beta \left( 1 - \frac{k^2+k'^2}{2}
%\right) \right) \right] \hat{a}_{\bf k'}^{(0)\dagger}
%\end{eqnarray}

\section{First Order Perturbation Theory}

The advantage of using Fourier transform lies in the fact, that the
equation (\ref{bofexp}) can be used in a natural way as a basis of
the perturbative expansion. Assuming a general solution of
$\delta(\K,t)$ in a perturbative form, $\delta(\K,t)=
\delta^{(0)}({\bf k},t)+\delta^{(1)}(\K,t)+\ldots$, we derive a
recurrent relation, which in the lowest order gives
% where $\lambda$ will be set to unity. The indexes $(i)$
%denote the order of the perturbative solution. Upon substituting Eq.
%(\ref{per_dec}) into Eq. (\ref{plain_evo}) we obtain
\begin{eqnarray}
&&i\partial_t\hat{\delta}^{(1)}(\K,t)= \frac{{\alpha}{\beta}^2}{8
\pi^{3/2}} e^{-t^2} \int \mbox{d}^3 k'\exp \left( -\frac{\beta^2 ({\bf k + k'})^2}{4}\right)\nonumber\\
&&\times\exp\left(-  i \beta t \left( 1 - \frac{k^2 +
k'^2}{2}\right) \right)  \hat\delta^{(0)\dagger} ({\bf
k'},t).\label{forder}
\end{eqnarray}
The first order equation (\ref{forder}) can be integrated and
consequently we can evaluate observable quantities like number of
scattered atoms or the correlation functions analytically.

\subsection{Density matrix}

%% The most straightforward observable quantity is the number of
%% elastically scattered atoms ejected from the condensate during the
%% collision.

In this subsection we find the approximate expression for the
density matrix of scattered atoms,
\begin{eqnarray}\label{rho}
&&\rho({\bf k_1,k_2},t) =  \langle \hat\delta^\dagger ({\bf k_1},t)
\hat\delta({\bf k_2},t)\rangle,
\end{eqnarray}
within the first order perturbation theory. We perform the
calculations in the momentum space.

Our system is initially in the vacuum state $|\Omega\rangle$, that
is $\hat\delta(\K, 0)|\Omega\rangle=0$. Hence, due to the
commutation relation,  $[\hat\delta ({\bf k},t) ,\hat\delta^\dagger
({\bf k'},t)]= \delta^3({\bf k - k'})$, at initial time we have
\begin{eqnarray}
\langle \hat\delta ({\bf k},0)
\hat\delta^\dagger ({\bf k'},0)\rangle= \delta^3({\bf k - k'}).\label{relation}
\end{eqnarray}
Then, upon substituting Eq.~(\ref{forder}) into (\ref{rho}), and
using Eq.~(\ref{relation}), we obtain
\begin{eqnarray} \label{rhocalka}
&&\rho({\bf k_1,k_2},t)= \frac{\alpha^2 \beta^4}{64 \pi^3} \int_0^t
\mbox{d} \tau \int_0^t\mbox{d} \tau'\exp \left( -\tau^2
-\tau'^2\right)\nonumber\\ \nonumber &&\times\exp\left(-
\frac{\beta^2 (k_1^2
+ k_2^2 )}{4}- i \beta (\tau k_1^2 - \tau' k_2^2)/2\right)\\
\nonumber &&\times \int \mbox{d}^3 k \exp \left(- \frac{\beta^2
(k^2 + {\bf k}( {\bf k_1 + k_2} ))}{2} \right)
\\
&& \times \exp \left(  i \beta (\tau -\tau') \left( 1-k^2/2 \right)
\right).
\end{eqnarray}
We show how to handle the above integral in Appendix A.  For $\beta
\gg 1 $ and $ \beta |{\bf k_1 + k_2}| \gg 1$ the density matrix
reduces to
\begin{eqnarray} \nonumber
&& \rho({\bf k_1,k_2},t) = \frac{\alpha^2 \beta \sqrt{2\pi}}{ 32
\pi^2 } \exp \left(- \frac{\beta^2 |{\bf k_1 - k_2} |^2}{8} \right)
\\ \nonumber
&& \times\int_0^t \mbox{d} \tau  \int_0^t \mbox{d} \tau' \, \exp
\left(- \frac{(\tau -\tau')^2|{\bf k_1 + k_2}|^2}{8} \right)
\\ \nonumber
&&\times\exp \left( -\tau^2 -\tau'^2+i \beta \tau  \left( 1 - \frac{|{\bf k_1 +
k_2}|^2}{8} -\frac{k_1^2}{2} \right) \right)\\
&&\times\exp \left( - i \beta \tau' \left( 1 - \frac{|{\bf k_1 +
k_2}|^2}{8} -\frac{k_2^2}{2} \right) \right).\label{rho_result}
\end{eqnarray}
This expression, although it still contains twofold integral over
time, is simple enough to give the distribution of scattered atoms,
and provide a further insight into coherence properties and dynamics
of scattering process.

\subsection{Final momentum distribution and coherence properties of
scattered atoms}
%at $t \rightarrow \infty$.}

\begin{figure}
\includegraphics[scale=0.315, angle=0]{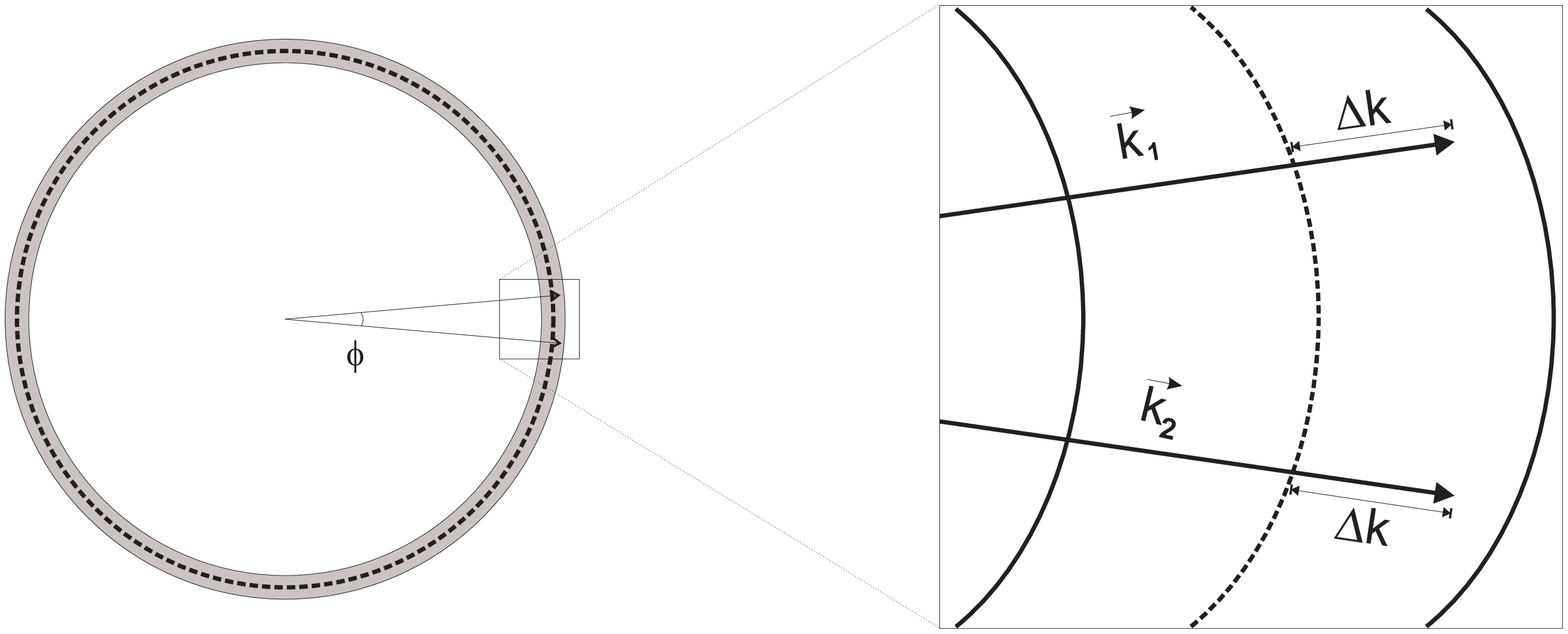}
\caption{Configuration of vectors $\K_1$ and $\K_2$ as in case a). The vectors are of the same length,
with angle $\phi$ between them. The dashed circle corresponds to $k=1$ and the grey area is the scattering
shell. The length of each of the vectors differs from unity by $\Delta k$.}\label{fig4}
\end{figure}

\begin{figure}
\includegraphics[scale=0.315, angle=0]{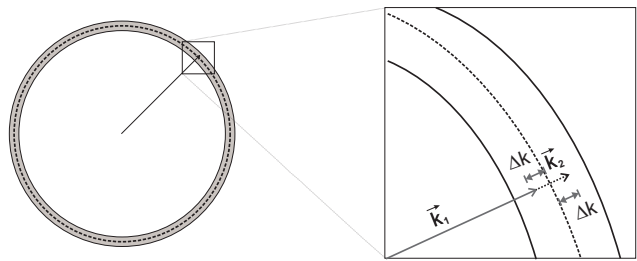}
\caption{Configuration of vectors $\K_1$ and $\K_2$ as in case b). The vectors are parallel.
The dashed circle corresponds to $k=1$ and the grey area is the scattering
shell. The lengths of $\K_1$ and $\K_2$ differ from unity by $\pm\Delta k$.}\label{fig5}
\end{figure}

From the last expression in the previous section
(Eq.~(\ref{rho_result})) the density matrix becomes negligible if
any of the arguments  $\beta |{\bf k_1 - k_2}|$, $\beta \left( 1 -
\frac{|{\bf k_1 + k_2}|^2}{8} -\frac{k_2^2}{2} \right) $ or $\beta
\left( 1 - \frac{|{\bf k_1 + k_2}|^2}{8} -\frac{k_1^2}{2} \right) $,
appearing in the exponential functions, are large (remember that we
work under constant assumption $\beta \gg 1$). Hence the only
regions where we expect nonvanishing value of $\rho({\bf
k_1,k_2},t)$ are those, for which wavevectors $k_1$ and $k_2$ are
close to each other ($|{\bf k_1 - k_2}| \ll 1 $), and they both have
the length close to unity (notice that in our units this coincides
with the length of the wavevector of the colliding condensates). It
is interesting to examine two cases:

a) Equal length case; $ k_1 = k_2 \equiv 1 + \Delta k $. In this
case (see Fig.~\ref{fig4}) the vectors $\K_1$ and $\K_2$ are of
equal length and form an angle $\phi$. The properties of the density
matrix strongly depend on the distance from the resonance surface
$\Delta k$. For small values of $\Delta k$ and small angles $\phi$
the density matrix can be approximated by the Gaussian function
\begin{eqnarray}\label{rho_width}
\rho (\Delta k,\phi)&=&\mathcal{A}\cdot \exp \left(- \frac{\beta^2
\phi^2}{8}  - \frac{\beta^2 \Delta k^2 }{2} \right).
\end{eqnarray}
The constant $\mathcal{A}=\frac{\alpha^2 \beta }{ 128 \sqrt{\pi}}
\left[ 1 + \frac{2}{\pi} \mbox{ArcTan}\left(\frac{1}{2
\sqrt{2}}\right) \right]$ was found evaluating (\ref{rho_result})
analytically on resonance ($\K_1=\K_2$ and $|\K_1|=1$).
Notice that  Eq.~(\ref{rho_width}) for $\phi=0$ represents the
density of the cloud of scattered atoms. The cloud has spherical
symmetry, following the symmetry of Eq.~(\ref{EofM}). It takes a
form of a shell localized around $k=1$ (which corresponds to
wave-vector $Q$ in our units). Its width is of order to $1/\beta$.

b) Parallel vectors case; ${\bf k_1} \parallel {\bf k_2}$ (see
Fig.\ref{fig5}). Wavevectors are shifted away from the resonance in
the opposite direction (${\bf k_1} = {\bf e} ( 1 -\Delta k )$ and
${\bf k_2} = {\bf e} (  1 +\Delta k) $). In this case we also find a
Gaussian fit
\begin{eqnarray} \nonumber
\rho (\Delta k)= \mathcal{A} \cdot \exp \left( - \frac{2}{3} \beta^2
\Delta k^2 + i \beta \Delta k\right).
\end{eqnarray}
This concludes our analysis of the density matrix. Now we move on to
the first order correlation function
\begin{equation}
 g_1(\K_1, \K_2, t)=\frac{\langle
 \hat\delta^\dagger(\K_1,t)\hat\delta(\K_2,t)\rangle} {\sqrt{\langle
 \hat\delta^\dagger(\K_1, t)\hat\delta(\K_1,
 t)\rangle\langle\hat\delta^\dagger(\K_2, t)\hat\delta(\K_2,
 t)\rangle}}.
\end{equation}
We find that in the case a)
\begin{eqnarray} \label{g11}
g_1(\Delta k,\phi) = \exp \left(- \frac{\beta^2 \phi^2}{8}  \right),
\end{eqnarray}
while in the case b)
\begin{eqnarray}\label{g12}
g_1 (\Delta k) =  \exp \left( - \frac{1}{6} \beta^2 \Delta k^2 + i
\beta \Delta k  \right).
\end{eqnarray}

\begin{figure}
\includegraphics[scale=0.18, angle=0]{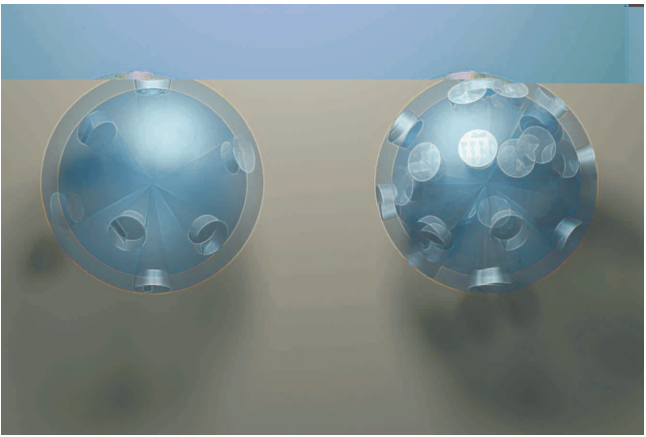}
\caption{The artistic representation of the atomic coherence. Final
states of scattered atoms lie on the spherical shell and the
coherence volume of each atom is marked as a dark region in this
shell. Case a) corresponds to weak scattering and in b) individual
coherence volumes start to overlap.}\label{fig9}
\end{figure}

By comparing Eq.~(\ref{g12}) describing the coherence in the radial
direction - $g_1 (\Delta k)$, and Eq.~(\ref{rho_width}) describing
the density $\rho (\Delta k,\phi=0)$ it seems reasonable to assume
that the radial coherence length of the shell of scattered atoms is
of the same order as its width. Following this observation we
present a very rough estimate of the number of atoms necessary for
the bosonic stimulation to occur. First we introduce the concept of
the coherence volume associated with each scattered atom. It is
determined by the first order correlation function. If we fix $\K_1$
and vary $\K_2$ we get a well defined peaked structure around
$\K_2=\K_1$. The size of this structure gives the coherence volume
associated with the single atom (see Fig.~\ref{fig9}). As we have
already shown, the radial extend of this structure is of order of
$1/\beta$. From Eq.~(\ref{g11}) we deduce that the angular coherence
length is of the same order $1/\beta$. Hence we estimate the
coherence volume to be equal to $1/\beta^3$. Next we remind
ourselves that the volume $V$ in the momentum space accessible by
the scattered atoms is spherical shell determined by the diagonal
part of the density matrix. In our case $V=4\pi/\beta$. When the
product of the number of scattered atoms times their coherence
volume is comparable with $V$, the coherence volumes associated with
distinct atoms start to overlap (see Fig.~\ref{fig9}). Then, the
wave nature of matter comes into play, and interference effects
would enhance the scattering - bosonic stimulation \cite{Feynmann}.
Using the above estimates, the critical number of atoms is equal to
$4 \pi \beta^2$. In the next section, we derive more practical
criterion for bosonic stimulation to show up in the collision of the
condensates.

\subsection{Number of scattered atoms for $t \gg 1/\beta $}

The most straightforward observable quantity, the number of
scattered atoms as a function of time can be expressed
in terms of the trace of the density matrix
\begin{eqnarray}
\label{number}\mathcal{S}(t)&=& \int d^3 k  \ \rho({\bf k,k},t).
\end{eqnarray}
Using Eq.~(\ref{rho_result}) we get:
\begin{displaymath}
\label{c1}{\mathcal S}^{(1)}(t)=\frac{\alpha^2 \beta}{4\sqrt{2\pi}}
\int_0^{\infty} k^2 \,\mbox{d} k \int_0^t \mbox{d} \tau \int_0^t
\mbox{d} \tau ' \, \exp \left(  - \tau^2 - \tau '^2\right)
\end{displaymath}
\begin{equation}
\times\exp\left( i  \beta (1-k^2)(\tau-\tau')- \frac{k^2}{2}(\tau -
\tau ')^2\right).
\end{equation}
Unless $k$ is close to unity the integral over $\tau$ and $\tau'$
will vanish due to the rapidly oscillating phase factor $\beta
(1-k^2)(\tau-\tau')$. Thus it is convenient to change the variables
$k = 1 +\frac{x}{\beta}$ and expand the integrand in the lowest
order of $x$ obtaining
\begin{eqnarray}
{\mathcal S}^{(1)}(t)&=&\frac{\alpha^2 \sqrt{2\pi}}{8 \pi}
\int_{-\infty}^{\infty} \mbox{d} x \int_0^t \mbox{d} \tau \int_0^t
\mbox{d} \tau '
\, \exp \left( - \tau^2 - \tau '^2  \right)\nonumber\\
&\times&\exp\left(i  2x (\tau-\tau') - \frac{1}{2}(\tau - \tau ')^2
\right).
\end{eqnarray}
Next, evaluating integral over $x$ we get
\begin{eqnarray}
{\mathcal S}^{(1)}(t)&=&\frac{\alpha^2 \sqrt{2\pi}}{8}
\int_0^t \mbox{d} \tau \int_0^t \mbox{d} \tau '\delta(\tau - \tau')\nonumber\\
&\times&\exp\left(- \tau^2 - \tau '^2- \frac{1}{2}(\tau - \tau ')^2
\right),
\end{eqnarray}
and finally the number of scattered atoms is equal to
\begin{eqnarray} \label{porow}
\mathcal{S}^{(1)}(t)= \frac{\pi \alpha^2}{16} \mbox{erf}(t
\sqrt{2}).
\end{eqnarray}
In Appendix B the same result is obtained within a classical model
of colliding hard spheres with cross-section equal to $8\pi a^2$.

\begin{figure}[htb]
\centering
\includegraphics[scale=0.31, angle=270]{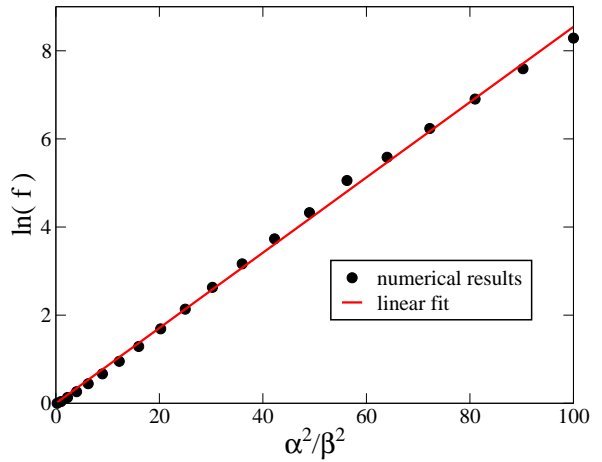}
\caption{Elastic collision loss as a function of $\alpha^2/\beta^2$.
The total number of scattered atoms is given by ${\mathcal
S}=\pi\frac{\alpha^2}{16}f\left(\frac{\alpha^2}{\beta^2}\right)$.
The figure shows that, $\ln(f)$ is a linear function of
$\alpha^2/\beta^2$ with slope equal to 0.085.} \label{fig8}
\end{figure}

An important remark concerns the condition for bosonic stimulation.
As we mentioned in the previous section it occurs when ${\mathcal
S}\geq4 \pi \beta^2$. By comparing this result with
Eq.~(\ref{porow}) (with $t \rightarrow \infty$), we get
$\alpha/\beta \geq 2$. This is a simple condition, which translated
into experimental parameters provides the criterion for bosonic
enhancement in the collision of two condensates. To verify this
condition we evaluate the total number of scattered atoms up to the
third order
\begin{equation}
{\mathcal S}^{(1)}+\mathcal{S}^{(2)}+\mathcal{S}^{(3)}
=\pi\frac{\alpha^2}{16}\left(1+c_1\frac{\alpha^2}
{\beta^2}+c_2\frac{\alpha^4}{\beta^4}\right)\label{third},
\end{equation}
where $c_1$ and $c_2$ are numerical coefficients. This suggests the
functional dependence of the form ${\mathcal
S}=\pi\frac{\alpha^2}{16}f\left(\frac{\alpha^2}{\beta^2}\right)$,
where $f(0)=1$. Our previous numerical results \cite{Zin}, obtained
both in perturbative and non-perturbative regimes confirm this
suggestion. In Fig.\ref{fig8} we plot the logarithm of $f$ versus
$(\alpha/\beta)^2$ and obtain a straight line with high accuracy.
Hence we see that indeed the bosonic stimulation (manifested by the
value of $f(\alpha^2/\beta^2)$ being larger than unity) occurs for
$\alpha/\beta \approx 2$. Moreover, function $f$ has an exponential
form $f[(\alpha/\beta)^2] \simeq \exp[0.085\,(\alpha/\beta)^2]$.
Using above formulas one can both easily estimate number of
scattered atoms and predict whether bosonic enhancement would be
present in the particular physical realization.

\subsection{Number of scattered atoms for $t \ll 1$. Characteristic
timescales of the signal build up.}

In order to analyze the dynamics of
$\langle\hat\delta^\dagger\hat\delta\rangle$ on a very short
timescale, we consider the diagonal part of the density matrix
\begin{eqnarray} \nonumber
&&\rho({\bf k,k},t)=\frac{\alpha^2 \beta \sqrt{2\pi}}{ 32 \pi^2 }
 \int_0^t \mbox{d} \tau \int_0^t \mbox{d} \tau' \, \exp
\left( -\tau^2 -\tau'^2 \right)
\\&&\times\exp \left( - \frac{(\tau -\tau')^2k^2}{2}  + i \beta (\tau
-\tau') \left( 1 - k^2 \right) \right).\label{jakis}
\end{eqnarray}
Notice that for very short times ($t \ll 1$), all the terms
proportional to squares of $\tau$ and $\tau'$ in the exponents
appearing in Eq.~(\ref{jakis}) can be dropped. With this
simplification the time integrals can be evaluated analytically,
giving
\begin{eqnarray} \label{short_time}
  \rho({\bf k,k},t) &=& \frac{\alpha^2 \beta \sqrt{2\pi}}{32 \pi^2}
  \frac{\sin^2 \left( \beta t(1-k^2)/2 \right) }{\left(  \beta (1-k^2)/2 \right)^2}.
\end{eqnarray}
\begin{figure}[htb]
\centering
\includegraphics[scale=0.3, angle=0]{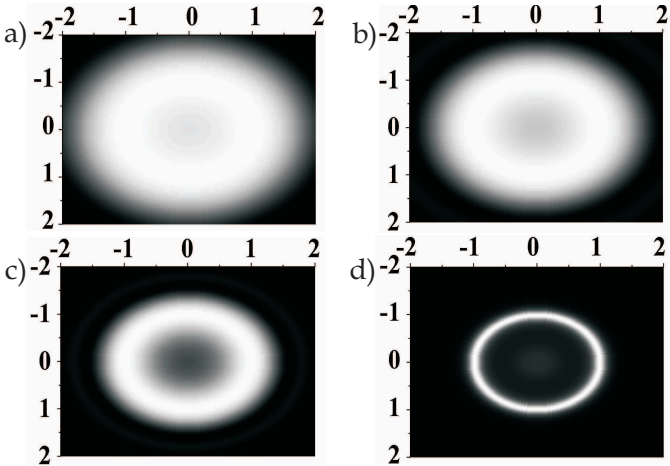}
\caption{The density of atoms in momentum space as given by
Eq.~(\ref{short_time}) for  $\alpha=10$ and $\beta=20$. The halo of
scattered atoms shrinks, forming a shell around $k=1$. The four
frames correspond to increasing times a) $t=0.07t_c$, b) $t=0.1t_c$,
c) $t=0.15t_c$ and b) $t=0.4t_c$, where $t_c$ is the collisional
time.} \label{fig6}
\end{figure}
The time evolution of the density of scattered atoms in $k_x$-$k_y$
plane is pictured in Fig.\ref{fig6}. It shows that atoms, which
initially scatter isotropically in the momentum space, eventually
form a well pronounced halo around the $k=1$.

To relate Eq.~(\ref{short_time})  to the number of scattered atoms,
we need to perform the integration over $\K$. Evaluating angular
integral explicitly and changing the variables $x = \beta t
(k^2-1)/2$, we get
\begin{eqnarray} \nonumber
 S(t) &=& \frac{\alpha^2 \sqrt{2\pi}}{8\pi } t \int_{-\beta t/2}^\infty
\mbox{d} x \, \frac{\sin^2 x}{x^2} \sqrt{\frac{2x}{\beta t}+1}.
\end{eqnarray}
This formula can be further simplified in the following two regimes:\\
a) $\beta t \ll 1$; we set the lower limit in the integral equal to
zero and use the approximation $ \sqrt{\frac{2x}{\beta t}+1} \simeq
\sqrt{\frac{2x}{\beta t}}$ to obtain
\begin{eqnarray} \nonumber
 S(t) &\simeq & \frac{\alpha^2 \sqrt{2\pi}}{8\pi } t \sqrt{\frac{2}{\beta t}}
  \int_{0}^\infty  \mbox{d} x \, \frac{\sin^2 x}{x^2} \sqrt{x} = \frac{\alpha^2
  }{4 \sqrt{\beta} } \sqrt{t}.
\end{eqnarray}
b) $ \beta t \gg 1$; we set the lower limit in the integral equal to
$-\infty$ and use the approximation $ \sqrt{\frac{2x}{\beta t}+1}
\simeq 1$ to obtain
\begin{eqnarray} \nonumber
S(t)&\simeq&\frac{\alpha^2\sqrt{2\pi}}{8\pi}t\int_{-\infty}^\infty\mbox{d}x\,
\frac{\sin^2 x}{x^2}=\frac{\alpha^2 \sqrt{2\pi}}{8}t.
\end{eqnarray}
Let's discuss the validity of the above result. Notice that we
introduced approximations describing two body interactions by a
contact potentials (renormalized delta function potential). On the
other hand any realistic potential should have a natural ultraviolet
cut-off $k_c$. If we extend the upper limit only up to $k_c$
(instead of extending it to $\infty$) we find in the case a)
discussed above a narrow window of very short times when the number
of scattered atoms grows quadratically in time. On later times
dynamics will not be affected by the cut-off. Consequently, the
realistic estimate would predict that the number of scattered atoms
should be quadratic function of time for initial extremely short
period, next there is a window of time when it behaves like a square
root, and finally it turns into linear, semiclassical, regime.

\subsection{Higher order correlation functions}

A lot of interesting information about the  quantum systems might be
obtained from its correlation functions. In this section we present
and discuss some of their properties. We focus on equal-time
correlation functions in momentum space. This choice is justified by
the fact that in most experiments performed with Bose Einstein
condensates, the common imaging technique is time-of-flight
absorption, which directly registers the momentum distribution of
the atomic cloud.

The normalized  $n$-th order density correlation function ($n \geq
2$) is defined as:
\begin{eqnarray}
&&g_n(\K_1,\ldots,\K_n, t)=\nonumber\\
&&=\frac{\langle\delta^\dagger(\K_1, t)\ldots\delta^\dagger(\K_n,
t)\delta(\K_n, t)\ldots\delta(\K_1, t)\rangle}
{\langle\delta^\dagger(\K_1, t)\delta(\K_1,
t)\rangle\ldots\langle\delta^\dagger(\K_n, t)\delta(\K_n,
t)\rangle}.
\end{eqnarray}
We make use of fact that the evolution of $\hat\delta$ is linear,
which means that $\hat\delta(\K,t)$ is a linear combination of
operators $\hat\delta(\K',0)$ and $\hat\delta^\dagger(\K',0)$; hence
\begin{eqnarray*}
\hat\delta(\K,t)=\int d{\K}'\left(U(\K,\K',t)\hat\delta(\K',0)+
V(\K,\K',t)\hat\delta^\dagger(\K',0)\right).
\end{eqnarray*}
For such a system Wick's theorem applies. Since the averages are
calculated in the vacuum state $|\Omega\rangle$, where
$\hat\delta(\K,0)|\Omega\rangle=0$ for all $\K$, all correlation
functions decompose into combination of products of a density matrix
$\rho(\K_i,\K_j, t)$ and anomalous density $m(\K_l,\K_m,t)$, defined
as
\begin{eqnarray}
m(\K_1,\K_2, t)=\langle\hat\delta(\K_1,t)\hat\delta(\K_2,t)\rangle.
\end{eqnarray}
This leads to the growing complexity of the correlation functions,
since in the $n$-th order, the number of terms contributing to the
$g_n$ is of the order of $(2n-1)!!$. For example, the second order
correlation function
\begin{eqnarray}
g_2(\K_1,\K_2, t)=\frac{\langle\hat\delta^\dagger(\K_1,
t)\hat\delta^\dagger(\K_2, t)\hat\delta(\K_2, t)\hat\delta(\K_1,
t)\rangle} {\langle\delta^\dagger(\K_1, t)\hat\delta(\K_1,
t)\rangle\langle\hat\delta^\dagger(\K_2, t)\hat\delta(\K_2,
t)\rangle}\nonumber
\end{eqnarray}
takes the form
%\begin{eqnarray}
%&&g_2(\K_1,\K_2, t)=\frac{\rho(\K_1,\K_1,t)\rho(\K_2,\K_2,t)+
%|\rho(\K_1,\K_2,t)|^2}{\rho(\K_1,\K_1,t)\rho(\K_2,\K_2,t)}\nonumber\\
%&&+\frac{|m(\K_1,\K_2,t)|^2}{\rho(\K_1,\K_1,t)\rho(\K_2,\K_2,t)}.
%\end{eqnarray}
\begin{eqnarray}
g_2(\K_1,\K_2, t)=1+\frac{|\rho(\K_1,\K_2,t)|^2+|m(\K_1,\K_2,t)|^2
}{\rho(\K_1,\K_1,t)\rho(\K_2,\K_2,t)}.\label{g2}
\end{eqnarray}
In the first order, we find an analytical expression for the
anomalous density
\begin{widetext}
\begin{eqnarray}\label{m}
&&m(\K_1,\K_2,t) = -i \frac{\alpha\beta^2}{8 \pi^{3/2}} \exp \left(
-\frac{\beta^2 ({\bf k_1 + k_2})^2}{4} \right) \int_0^t \mbox{d}
\tau \,  \exp \left( -\tau^2 -  i \beta \tau \left( 1 - \frac{k_1^2
+ k_2^2}{2}\right) \right)
\\ \nonumber \\ \nonumber
&&= -i   \frac{\alpha \beta^2}{16 \pi } \exp \left( - \frac{\beta^2
({\bf k_1} +{\bf k_2})^2}{4} - \frac{\beta^2 }{4}
\left(1-\frac{k_1^2+k_2^2}{2} \right)^2\right) \left[\mbox{erf}
\left( t + \frac{i}{2} \beta\left( 1 - \frac{k_1^2+k_2^2}{2} \right)
\right) - \mbox{erf}\left( \frac{i}{2} \beta \left( 1 -
\frac{k_1^2+k_2^2}{2} \right) \right)  \right].
\end{eqnarray}
\end{widetext}
The above function reaches its maximum when $\K_1 = -\K_2$ and both
vectors are of unit length ($k_1 =k_2 =1$) due to the energy and
momentum conservation.

\begin{figure}[htb]
\centering
\includegraphics[scale=0.35, angle=270]{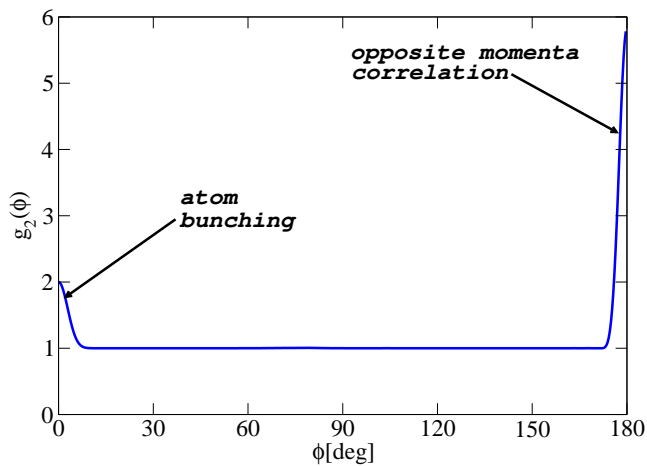}
\caption{The second order correlation function for $k_1=k_2=1$ as a
function $\phi$ --  relative angle between the vectors and
$\alpha=20$, $\beta=30$. The strong maximum around $\phi=0$
corresponds to atom bunching. The maximum around $\phi=180^\circ$ is
an effect of correlations of pairs of atoms with opposite momenta.}
\label{fig10}
\end{figure}

Combining Eqs~(\ref{g2}) and ~(\ref{m}) with the expressions for the
density matrix, discussed before, we can find analytic expression
for the second order correlation function in the first order
perturbation theory. In Fig.\ref{fig10} we show the second order
correlation function for $k_1=k_2=1$ as a function of relative angle
$\phi$ between the vectors. The maximum around $\phi=0$ corresponds
to the detection of two scattered particles close to each other. It
is greater than one due to the bosonic statistics of scattered
atoms. This effect is called {\it atom bunching}. Since the atoms
scatter in pairs of opposite momenta, the $g_2$ correlation function
reveals strong correlation around $\phi=180^\circ$. It is
exclusively an anomalous density part of the second order
correlation function that contributes to this maximum.

\section{Conclusions}

We studied the nature of the transition from spontaneous to
stimulated scattering regime in the collision of two Bose Einstein
condensates. Within the first order approximation we derived
analytical expressions for the density matrix and anomalous density.
This enabled us to derive the condition for bosonic stimulation
effect. The characteristic timescales in the dynamics of scattered
atoms were identified. Finally, we showed that scattered atoms
feature both bunching as well as the opposite momenta correlations.

\section{Acknowledgement}

The authors would like to acknowledge support from KBN Grant
2P03B4325 (P.Z. and J. C.) and the Polish Ministry of Scientific
Research and Information Technology under grant PBZ-MIN-008/P03/2003
(M.T.). Special thanks are addressed to Prof. Kazimierz
Rz\c{a}\.zewski for valuable and stimulating discussions and to
Baltazar Brukalski for creating artistic view presented in
Fig.~\ref{fig9}.

\section{Appendix A}

In this appendix we simplify the expression for density matrix
(\ref{rhocalka}) upon evaluating the integral over $\mbox{d}^3k$ in
the limit $\beta \gg 1$. The angular part of this integral can be
performed analytically
\begin{eqnarray}
&&\int \mbox{d}^3 k\,\exp\left(- \frac{\beta^2  (k^2 + {\bf k}( {\bf
k_1 + k_2} ))}{2}\right)
\nonumber\\
&&\times\exp\left(i\beta (\tau -\tau') \left( 1-\frac{k^2}{2} \right) \right)=\nonumber\\
&&2 \pi \int_0^\infty k^2 \mbox{d} k \, \frac{4}{\beta^2 |{\bf k_1 +
k_2}| k}   \sinh \left(
\frac{\beta^2 k |{\bf k_1 + k_2} |}{2} \right)\nonumber\\
&&\times\exp\left(-\frac{\beta^2k^2}{2}+i\beta (\tau -\tau') \left(
1-\frac{k^2}{2} \right) \right). \nonumber
\end{eqnarray}
Next, using the identity $2(k_1^2+k_2^2) = |{\bf k_1 + k_2}|^2 +
|{\bf k_1 - k_2}|^2$, we get
\begin{widetext}
\begin{eqnarray} \nonumber
&&\rho({\bf k_1,k_2},t)=\frac{\alpha^2\beta^2}{ 16 \pi^2  } \frac{1}{ |{\bf k_1 + k_2}|}
\exp \left(- \frac{\beta^2 |{\bf k_1 - k_2} |^2}{8}  \right)
\int_0^t \mbox{d} \tau  \int_0^t \mbox{d} \tau' \, \exp \left(
-\tau^2 -\tau'^2 - i \beta (\tau k_1^2 - \tau' k_2^2)/2 + i \beta
(\tau -\tau')  \right)
\\ \label{g1}
&&\times\int_0^\infty k \mbox{d} k\exp \left(-i \frac{\beta}{2}
(\tau -\tau') k^2 \right) \left[ \exp \left( - \frac{\beta^2}{2}
\left(k-\frac{|{\bf k_1 + k_2} |}{2} \right)^2\right) - \exp \left(
- \frac{\beta^2}{2} \left(k+\frac{|{\bf k_1 + k_2} |}{2} \right)^2
\right)  \right].
\end{eqnarray}
\end{widetext}
The most significant contribution to the above integral comes from
the region, where the condition $ \beta |{\bf k_1 + k_2}| \gg 1$ is
satisfied. In this region, we can neglect
$\exp\left(-\frac{\beta^2}{2}\left(k+\frac{|{\bf
k_1+k_2}|}{2}\right)^2\right)$ in comparison with $\exp \left( -
\frac{\beta^2}{2} \left(k-\frac{|{\bf k_1 + k_2} |}{2}
\right)^2\right)$. Additionally we introduce a new variable $ k =
\frac{|{\bf k_1 + k_2}|}{2} + \frac{x}{\beta} $ and get
\begin{widetext}
\begin{eqnarray} \label{ar}
& &\int_0^\infty k \mbox{d} k \,  \exp \left(-i \frac{\beta}{2}
(\tau -\tau') k^2 \right)\exp \left( - \frac{\beta^2}{2}
\left(k-\frac{|{\bf k_1 + k_2} |}{2} \right)^2 \right)
\\ \nonumber
& & = \frac{1}{\beta^2} \int_{-\beta |{\bf k_1 + k_2}|/2 }^\infty
\mbox{d} x \ \left(\beta |{\bf k_1 + k_2}|/2 + x \right) \exp
\left(-\frac{i}{8} (\tau -\tau')\left( \beta |{\bf k_1 + k_2}|^2 + 4
x |{\bf k_1 + k_2}| + \frac{4x^2}{\beta} \right) \right)\exp \left(
- x^2 /2 \right).
\end{eqnarray}
\end{widetext}
The condition $ \beta |{\bf k_1 + k_2}| \gg 1$, together with the
presence of the exponential factor $\exp \left( - x^2 /2 \right)$,
allow to extend the lower limit in the integral over $x$ in
Eq.~(\ref{ar}) to $-\infty$, and neglect $x$ in comparison with
$\beta |{\bf k_1 + k_2}|/2$. Due to the presence of the exponential
factor $\exp \left( -\tau^2 - \tau'^2 \right)$ in the equation
(\ref{g1}) we can make another approximation, namely neglect the
phase factor $i(\tau - \tau')\frac{x^2}{2\beta}$. Having all above
approximation applied we can evaluate simple Gaussian integral to
get finally
\begin{widetext}
\begin{eqnarray} \nonumber
&&\rho({\bf k_1,k_2},t)=\frac{\alpha^2\beta \sqrt{2\pi}}{ 32 \pi^2}
\exp \left(- \frac{\beta^2 |{\bf k_1 - k_2} |^2}{8} \right) \int_0^t
\mbox{d} \tau  \int_0^t \mbox{d} \tau' \, \exp \left( -\tau^2
-\tau'^2  - \frac{(\tau -\tau')^2|{\bf k_1 + k_2}|^2}{8} \right)
\\ \label{density}
&&\times\exp \left( i \beta \tau  \left( 1 - \frac{|{\bf k_1 + k_2}|^2}{8}
-\frac{k_1^2}{2} \right) - i \beta \tau' \left( 1 - \frac{|{\bf k_1
+ k_2}|^2}{8} -\frac{k_2^2}{2} \right) \right).
\end{eqnarray}
\end{widetext}

\section{Appendix B}

In this appendix we calculate the collisional loss for two
counter-propagating clouds of classical particles. This model is
meant as a classical counterpart of the collision described in the
main body of the paper. To match the conditions used in the paper we
assume a Gaussian density profile for each of the clouds and we
restrict ourself to the dilute gas limit ($n \sigma_0 \sigma \ll
1$), where $\sigma_0$ is a cross-section for a collision of two
particles and $\sigma$ is the radius of the cloud. In this limit we
neglect secondary collisions. We also ignore the depletion of the
clouds; this approximation is valid as long as the fraction of
scattered atoms is small. When the  collision is described in the
reference frame associated with one of the clouds, the density of
this cloud is equal to
\begin{eqnarray}
n_1(\x)=\frac{N}{2\pi^{3/2}\sigma^{3}}
\exp\left[-\frac{x_1^2+x_2^2+x_3^2}{\sigma^2}\right], \nonumber
\end{eqnarray}
while the other cloud propagates along the $x_1$ axis with
velocity $2v$,
\begin{eqnarray}
n_2(\x, t)=\frac{N}{2\pi^{3/2}\sigma^{3}}
\exp\left[-\frac{(x_1+2vt)^2+x_2^2+x_3^2}{\sigma^2}\right].
\nonumber
\end{eqnarray}
Here both clouds contain $N/2$ particles.

The number of particles scattered up to the time $t$ equals
\begin{eqnarray}
{\mathcal S}(t) = 2\times\left[\int \mbox{d}^3 r\ n_1(\x) \int_0^t
\mbox{d}t'\ \left(n_2(\x,t')\cdot 2v\sigma_0\right)\right].\nonumber
\end{eqnarray}
Here, the prefactor ``2'' on the right-hand side accounts for the
fact that in every collision two particles are scattered. For two
identical bosons, the cross-section $\sigma_0$ at low-energy limit
is equal to $8\pi a^2$, where $a$ is the s-wave scattering length.
Using the identity
\begin{eqnarray*}
\int_{-\infty}^\infty\exp\left(-\frac{x^2}{\sigma^2}\right)
\mbox{erf}\left(\frac{x+2vt}{\sigma}\right){\mathrm d}x=
\sqrt{\pi}\sigma\cdot\mbox{erf}\left(\frac{\sqrt{2}vt}{\sigma}\right)\nonumber
\end{eqnarray*}
and rescaling time ($t/t_C \rightarrow t$), where the collisional
time $t_C=\sigma/v$, we obtain:
\begin{eqnarray}
{\mathcal S}(t) = \frac{N^2 a^2}{\sigma^2} \mbox{erf}(t\sqrt{2})=\frac{\alpha^2\pi}{16}\mbox{erf}(t\sqrt{2}).\nonumber
\end{eqnarray}
This result reproduces number of scattered atoms obtained in the
quantum model within the first order perturbation theory.


\begin{thebibliography}{30}

\bibitem{Ovchinnikov99}
Yu.  B. Ovchinnikov {\it et al.}, Phys. Rev. Lett. {\bf 83}, 284
(1999).
% Yu.  B. Ovchinnikov, J. H. M\"{u}ller, M. R. Doery, E. J. D.
% Vredenbregt, K. Helmerson, S. L. Rolston, and W. D. Phillips, Phys.
% Rev. Lett. {\bf 83}, 284 (1999).

\bibitem{Kozuma99}
M. Kozuma {\it et al.}, Phys.  Rev.  Lett.  {\bf 82}, 871 (1999).
% M. Kozuma, L. Deng, E. W. Hagley, J. Wen, R. Lutwak, K. Helmerson, S.
% L. Rolston, and W. D. Phillips, Phys.  Rev.  Lett.  {\bf 82}, 871
% (1999).

\bibitem{Stenger}
J. Stenger {\it et al.}, Phys.  Rev.  Lett. {\bf 82}, 4569 (1999).
% J. Stenger, S. Inouye, A. P. Chikkatur, D. M. Stamper-Kurn, D. E.
% Pritchard, and W. Ketterle, Phys.  Rev.  Lett. {\bf 82}, 4569 (1999).

\bibitem{nist_al}
E. W. Hagley {\it et al.}, Science {\bf 283}, 1706 (1999).
% E. W. Hagley, L. Deng, M. Kozuma, J. Wen, K. Helmerson, S. L. Rolston,
% and W. D. Phillips, Science {\bf 283}, 1706 (1999).

\bibitem{nist_cl_exp}
E.W. Hagley {\it et al.}, Phys. Rev. Lett. {\bf 83}, 3112 (1999).
% E.W. Hagley, L. Deng, M. Kozuma, M. Trippenbach, Y.\ B.\ Band,
% M. Edwards, M. Doery, P.\ S.\ Julienne, K. Helmerson, S.\ L.\ Rolston,
% and W.\ D.\ Phillips, Phys. Rev. Lett. {\bf 83}, 3112 (1999).

\bibitem{cl_theory}
M. Trippenbach {\it et al.}, J. Phys. B {\bf 33}, 47-54 (2000).

\bibitem{Meystre} G. Lenz, P. Meystre, and E. W. Wright, Phys. Rev. Lett. {\bf 71},
3271 (1993), E. Goldstein, K. Plattner, and P. Meystre, Quantum
Semiclassic. Opt. {\bf 7}, 743 (1995), E. Goldstein, K. Plattner,
and P. Meystre, J. Res. Natl. Inst. Stand. Technol. {\bf 101}, 583
(1996), E. Goldstein and P. Meystre, Phys. Rev. A {\bf 59}, 1509
(1999).

\bibitem{4WM_1}
M. Trippenbach, Y. B. Band, and P. S. Julienne, Optics Express {\bf
3}, 530 (1998),
%Theory of four-wave mixing of matter waves from a
%Bose-Einstein condensate

\bibitem{Tripp} M. Trippenbach, Y. B. Band,1 and P. S. Julienne, Phys. Rev. A
{\bf62}, 023608 (2000).

\bibitem{4WM_2}
L. Deng {\it et al.}, Nature {\bf 398}, 218-220 (1999).

\bibitem{johny} J. M. Vogels, K. Xu, and W. Ketterle,  Phys. Rev. Lett.
{\bf 89}, 020401 (2002)

\bibitem{Katz} N. Katz, R. Ozeri, E. Rowen, E. Gershnabel, and N.
Davidson, Phys. Rev. A 70, 033615 (2004);
%Decoherence and dephasing in strongly driven colliding Bose-Einstein condensates
N. Katz, E. Rowen, R. Ozeri, and N. Davidson, cond-mat/0505762.
%Collisional decay of a strongly driven Bose-Einstein condensate

\bibitem{Band} Y. B. Band {\it et al.}, Phys. Rev. Lett. {\bf 84}, 5462 (2000)

\bibitem{Chwed} J. Chwede\'{n}czuk, M. Trippenbach and K. Rz\c{a}\.{z}ewski,
J. Phys. B, L391, {\bf 37} (2004)

\bibitem{norrie} A.A. Norrie, R.J. Ballagh and
C.W. Gardiner, Phys. Rev. Lett. {\bf 94}, 040401 (2005), see also A.
Sinatra, Y. Castin and C. Lobo, Journal of Modern Optics {\bf 47},
2629 (2000)

\bibitem{Bach} R. Bach, M. Trippenbach and K. Rz\c{a}\.zewski,
Phys. Rev. A {\bf 65} 063605 (2002)

\bibitem{Yuro} V. A. Yurovsky, Phys. Rev. A {\bf 65} 033605 (2002)

\bibitem{Zin} P. Zi\'{n} {\it et al.}, Phys. Rev. Lett. {\bf 94}, 200401 (2005)

\bibitem{CDG}
%Bose-Einstein Condensates in Time Dependent Traps
Y. Castin and R. Dum,  Phys. Rev. Lett. {\bf 77}, 5315-5319 (1996);
%Evolution of a Bose gas in anisotropic time-dependent traps
Yu. Kagan, E. L. Surkov, and G. V. Shlyapnikov,  Phys. Rev. A {\bf
55}, R18-R21 (1997)

\bibitem{Abram} M. Abramovich, I. A. Stegun, {\it "Handbook of Mathematical
Functions With Formulas, Graphs and Mathematical Tables"},  Dover
Publications 1974.

\bibitem{Feynmann} R. P. Feynman, {\it "Feynman Lectures On Physics", Volume
3}, Addison Wesley Longman (1970).

\bibitem{Mandel} L. Mandel and E. Wolf,{\it "Optical Coherence and Quantum Optics"},
Cambridge (1995).




\end{thebibliography}
\end{document}